\documentclass{emulateapj}

\slugcomment{Accepted for publication in the Astrophysical Journal Letters}

\shorttitle{Optical Bursts from MS\,1603.6+2600}
\shortauthors{Hynes et al.}

\begin{document}


\title{Discovery of Optical Bursts from MS\,1603.6+2600 = UW~CrB}

\author{R. I. Hynes\altaffilmark{1},
	E. L. Robinson, E. Jeffery}

\affil{McDonald Observatory and Department of Astronomy, 
The University of Texas at Austin, 
1 University Station C1400, Austin, Texas 78712}
\altaffiltext{1}{Hubble Fellow; rih@astro.as.utexas.edu}

\begin{abstract}
We report the discovery of several optical burst-like events from the
low-mass X-ray binary MS~1603.6+2600 (UW~CrB).  The events last for a
few tens of seconds, exhibit a very fast rise and slow decay, and
involve optical brightening of a factor of 2--3.  The flares appear
distinct from the lower level flickering and instead strongly resemble
reprocessed type-I X-ray bursts as seen in a number of other neutron
star low-mass X-ray binaries.  In conjunction with the previously
reported candidate X-ray burst, these confirm that the compact object
in UW~CrB is a neutron star.  We examine the optical burst brightness
and recurrence times and discuss how the nature of the system can be
constrained.  We conclude that the source is most likely an accretion
disk corona source at an intermediate distance, rather than a nearby
quiescent system or very distant dipper.
\end{abstract}


\section{Introduction}
\label{IntroSection}
The X-ray source MS\,1603.6+2600 was discovered in the Einstein
Extended Medium Sensitivity Survey \citep{Gioia:1990a} and associated
with a faint ($R=19.4$) optical counterpart designated UW~CrB
\citep{Morris:1990a}.  Its nature has remained a puzzle.
\citet{Morris:1990a} found the counterpart to be an eclipsing binary
with an orbital period of 111.04\,min and considered the source to be
either a cataclysmic variable or low-mass X-ray binary (LMXB) hosting
a neutron star.  The emission line spectrum and optical to X-ray flux
ratio favored the LMXB interpretation, with an accretion disk corona
(ADC) source most likely.  The implied distance was large,
30--80\,kpc, making this high latitude source a halo object.
\citet{Hakala:1998a} reconsidered these possibilities and proposed
another alternative -- a quiescent low-mass X-ray binary, likely a
black hole system, which is much closer to us.  An important clue was
subsequently provided by \citet{Mukai:2001a} who identified a strong
X-ray flare in {\it ASCA} data.  While this appeared to resemble a
type-I X-ray burst, the authors did not consider this identification
conclusive.  If the event was a type-I burst then it was faint,
indicating either a very distant object in the halo, or an ADC source.
Based on the X-ray lightcurve, \citet{Mukai:2001a} favored the former
of these interpretations, arguing that the source is a dipper rather
than an ADC source.  Finally, \citet{Jonker:2003a} reported new {\it
Chandra} observations of the source, and favored the ADC
interpretation, although they allowed that a quiescent system was
still possible if the earlier X-ray flare was not a type-I burst.
They rejected the distant dipper scenario, arguing that the optical
luminosity would then be too high for a compact 2\,hr binary.

From a more theoretical standpoint, \citet{Ergma:1993a} considered
several evolutionary scenarios for the LMXB case, including degenerate
and non-degenerate hydrogen rich mass donors and evolved helium stars.
Again, bursts could be a crucial diagnostic.  The presence, recurrence
time, and duration of bursts can discriminate between systems with
different mass transfer rates (e.g., the degenerate and non-degenerate
cases discussed by \citealt{Ergma:1993a}), and the burst properties
will be sensitive to the chemical composition of the accreted
material.

To date, the only published X-ray burst from this source
was that reported by \citet{Mukai:2001a}, and this only yielded
60\,counts.  It is possible to also search for bursts in the optical,
as the optical counterpart, while faint, is accessible to rapid
photometry.  Type-I X-ray bursts are expected to be manifested in the
optical via reprocessed X-ray emission.  This behavior has been widely
seen in many other LMXBs for several decades (e.g.,
\citealt{Grindlay:1978a} and many subsequent works).  Optical bursts
are dramatic, involving a brightening of a factor $\sim 2$.
Ultraviolet bursts are also present and are even more dramatic
\citep{Hynes:2004a}.

We report here rapid optical photometry of UW~CrB.  The primary goal
of the program was to resolve the flickering contamination of the
orbital variability, and this study will be presented separately.
However, several optical bursts were serendipitously discovered, and
we discuss those here.

\section{Observations}
UW~CrB was observed over several nights from 2004 April 16--26 using
the Argos fast CCD camera \citep{Nather:2004a} on the McDonald
Observatory 2.1\,m telescope.  A total of 25\,hrs of good data were
obtained; details are provided in Table~\ref{LogTable}.  Observations
were obtained with a broadband $BVR$ filter (Rolyn Optics No.\
66.2475, bandpass $\sim4000-7500$\,\AA) to maximize count rates and
hence allow exposure times of 5 or 10\,s.  The data were taken as a
continuous sequence of images with negligible intervening dead-time.
Conditions were mostly near-photometric with 1--2\,arcsec seeing and
no Moon.  The nights of April 18, 20, and 26 experienced poorer
transparency and/or seeing.

\begin{table}
\caption{Log of observations}
\label{LogTable}
\begin{tabular}{lcccc}
\hline
\noalign{\smallskip}
Date & UT Range & Good          & Exposure & Bursts \\
(2004)     &          & time (ks)\tablenotemark{a} & time (s) &     \\
\noalign{\smallskip}
\hline
\noalign{\smallskip}
April 16 & 08:04:21--11:43:11 & 11.4 & 10 & 1 \\
April 18 & 10:14:53--11:29:03 &  4.1 & 10 & 0 \\
April 20 & 06:28:40--11:25:55 &  5.8\tablenotemark{b} & 10 & 0 \\
April 21 & 08:01:05--11:37:65 & 13.0 & 10 & 1 \\
April 22 & 07:39:13--11:26:33 & 13.6 &  5 & 1 \\
April 23 & 05:43:09--11:31:19 & 20.9 &  5 & 1 \\
April 25 & 09:20:34--11:40:34 &  8.0 &  5 & 0 \\
April 26 & 05:35:11--06:58:11 &  4.8 &  5 & 0 \\ 
         & 06:58:26--10:45:36 &  9.0\tablenotemark{b} & 10 & 0 \\
\noalign{\smallskip}
\hline
\tablenotetext{a}{Total period when reliable differential photometry
  could be obtained, not always uninterrupted.}
\tablenotetext{b}{These long runs occurred in very poor conditions and
  limited interrupted good intervals occurred; some bursts may have been missed.}
\end{tabular}
\end{table}

Data reduction employed a combination of {\sc iraf} routines to
generate calibration files and then a custom IDL pipeline to apply
calibrations and extract photometry.  Bias structure and dark current
were subtracted using many dark exposures of the same duration as the
object frames.  Residual time-dependent bias variations were removed
using two partial bad columns which are not light sensitive.
Sensitivity variations were removed using flat-field exposures of the
inside of the dome; we verified that this flattened the average sky
background and hence that no illumination correction was necessary.

Photometry was extracted using standard aperture photometry
techniques.  For each lightcurve, the aperture was chosen to minimize
the scatter between stars C and V of \citet{Hakala:1998a}, and
differential photometry of UW~CrB was then performed relative to star C.
Adopted apertures were typically 1.0--1.5\,arcsec in radius.

\section{Lightcurves}

\begin{figure}
\includegraphics[scale=0.35]{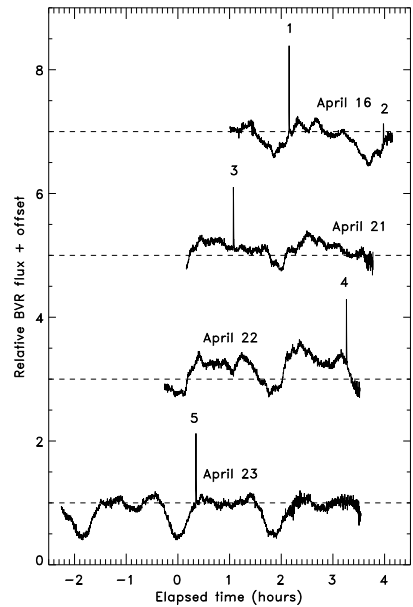}
\caption{Selected lightcurves of UW~CrB.  Only those exhibiting bursts
  are shown.  All
  lightcurves have been normalized relative to the mean flux level
  observed during the run.  Offsets have been applied to avoid overlap; dashed
  lines indicate the unit flux level for each curve.  Lightcurves have
  been shifted by an integral number of orbital periods to bring
  eclipses into alignment, but no attempt has been made to absolutely
  phase them.  5\,s lightcurves have been rebinned to 10\,s resolution
  for clarity and consistency.  Numbers indicate bursts.}
\label{OrbitFig}
\end{figure}

We show the lightcurves containing bursts in Fig.~\ref{OrbitFig}.  These
exhibit a typical range of morphologies for this source
\citep{Hakala:1998a}, with eclipses sometimes being very deep, and
sometimes barely detectable.  Superposed on the orbital modulations
and low-level flickering are several very sharp, large amplitude
burst-like events, involving a flux increase of a factor of 2--3.  We
will discuss their origin in Section~\ref{DiscussionSection}, but here
we will anticipate the conclusion by characterizing them in the same
terms commonly used for type-I X-ray bursts.

The events appear to be real, being resolved in time and exhibiting no
anomalies in the point spread function (PSF).  Event 2 is of lower
brightness than the others, but also appears resolved in time and has
a normal PSF.

\begin{figure}
\includegraphics[scale=0.35]{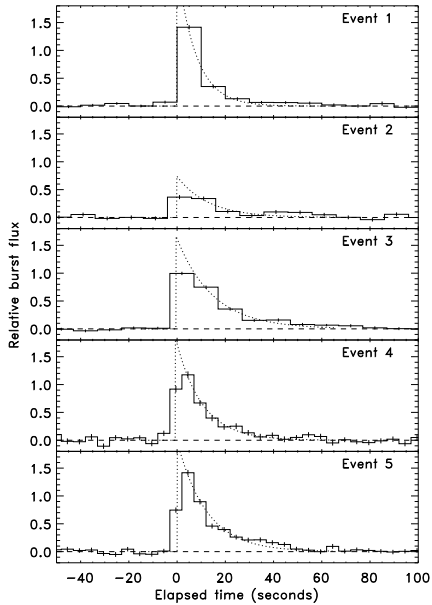}
\caption{Expanded view of each candidate type-I burst.  Histograms
  indicate the data with statistical error-bars.  The dotted line is
  an exponential decay fit to each burst.  As in Fig.~\ref{OrbitFig},
  fluxes are relative to the source mean brightness and so 
can be directly compared.  The persistent level has been subtracted
  off using fluxes immediately before and after each burst.}
\label{BurstFig}
\end{figure}

To characterize the bursts we fit a simple model consisting of an
instantaneous rise and exponential decay.  As the burst peak is
unresolved in the data, we rebin this model to match the data in
fitting.  Free parameters are the burst amplitude, start-time, and
e-folding decay time.  Expanded views of each burst are shown in
Fig.~\ref{BurstFig} together with the best fitting models.  Results
are summarized in Table~\ref{BurstFitTable}, based on fitting data
from $-20$\,s to $+60$\,s.  We quote the relative fluence, expressed
as individual burst fluence values divided by the mean fluence of
bursts 1, 3, 4, and 5, and the burst durations $\tau$, defined by
\citet{vanParadijs:1988a} as the burst fluence divided by the peak
flux.  The latter is obviously crudely constrained as the peaks are
not actually resolved.  For both quantities we quote both values from
the model fit and those derived directly from the data.  The relative
fluences derived are similar with both methods.  The durations differ
significantly, however.  Durations derived from the data probably
overestimate the burst duration, as the burst peak is unresolved and
hence the peak flux will be underestimated.  In contrast, the model
assumes a sharper peak than is seen in type-I bursts, so will
underestimate the duration compared to such bursts.  Hence the two
values obtained for each burst should bracket its true duration.
Given that we do not resolve the peak, this is probably the best
estimate that can be made.

\begin{table}
\caption{Properties of burst fits}
\label{BurstFitTable}
\begin{tabular}{lccccc}
\hline
\noalign{\smallskip}
Burst & \multicolumn{2}{c}{Relative fluence\tablenotemark{a}} &
& \multicolumn{2}{c}{Duration (s)} \\
\cline{2-3}\cline{5-6}
      & Model & Observed & & Model\tablenotemark{b} & Observed \\
\noalign{\smallskip}
\hline
\noalign{\smallskip}
1     & 0.91 & 0.93 &&  7.8 & 15.2 \\ 
2     & 0.43 & 0.45 && 13.6 & 28.9 \\
3     & 1.11 & 1.09 && 15.0 & 25.3 \\
4     & 0.90 & 0.92 && 10.7 & 18.2 \\
5     & 1.08 & 1.06 && 12.1 & 17.2 \\
\noalign{\smallskip}
\hline
\tablenotetext{a}{Relative to the mean of bursts 2, 4, 5, and 6.}
\tablenotetext{b}{For the exponential decay model, the duration is
  equal to the e-folding time.}
\end{tabular}
\end{table}

The bursts are, with the exception of the weak event 2, very uniform
in properties.  The burst fluence is constant to within 10\,\%, and
burst durations (including event 2) are all consistent with a value of
5--15\,s if measured from the model (probably an underestimate), or
15--30\,s from the unresolved data (an overestimate).  ``True'' values
of the burst duration (in the sense of \citealt{vanParadijs:1988a})
are likely to be in the 10--20\,s range for these bursts (for example,
events 4 and 5 which are better resolved).

Burst recurrence times can be crudely estimated from the data,
although are subject to small number statistics and sampling problems.
We observed four full bursts in 90.6\,ks (25.2\,hrs) of good data, or
an average of 1 per 6.3\,hrs.  If the bursts occur randomly (i.e., as
Poisson distributed events) then recurrence times of between 3.8 and
11\,hrs have a 10\,\%\ or greater chance of producing 4 events in this
period (i.e., longer or shorter recurrence times would be excluded at
90\,\% confidence).  Of course, if the events are type-I X-ray bursts
then they are not distributed as Poisson events but occur
quasi-regularly.  In that case, the statistics could be biased by our
once-per-day sampling.  More robust constraints are that the longest
period continuously observed without a burst being seen is 3.5\,hrs,
and the shortest period between {\em observed} bursts is 21.1\,hrs.
Both arguments suggest that an average recurrence time of much less
than 4\,hrs is very unlikely.  Recurrence times of up to $\sim12$\,hrs
are possible, or around 24\,hrs if they were synchronized with our
observations, and irregular enough to allow intervals as short as
21.1\,hrs.

\section{Discussion}
\label{DiscussionSection}

\subsection{Comparison with type-I bursts}

It has already been suggested by \citet{Mukai:2001a} that a type-I
X-ray burst was seen by {\it ASCA}.  The optical events that we see
also exhibit characteristics typical of type-I bursts.  The implied
duration of $\tau\sim 10-20$\,s, the fast rise and slow decay, and
optical brightening of a factor $\sim2-3$ are all similar to
reprocessed optical bursts seen in other objects (e.g.,
\citealt{Grindlay:1978a}; \citealt{Schoembs:1990a};
\citealt{Robinson:1997a}; \citealt{Homer:1998a}; and other works).
The burst frequency, one per $\sim4.2$\,hrs, is also normal.

Event 2 breaks the pattern of otherwise uniform burst fluences, but
has a similar duration to the other events.  It is also the only burst
seen in the same lightcurve as another, suggesting that it could be a
mini-burst produced when the first burst (event 1) leaves unspent fuel
on the neutron star surface.  Such behavior is seen in other LMXBs
(e.g., \citealt{Gottwald:1986a}).

The uniformity of fluences also presents suggestive, if not
conclusive, evidence for the location of reprocessing of the bursts.
Since UW~CrB is a high inclination system, if the burst reprocessing
were dominated by the inner face of the companion star, we would
expect large changes in the brightness of optical bursts dependent on
orbital phase.  In particular, event 3 occurs near phase 0.5 when the
inner face of the companion is viewed nearly face-on, hence should be
stronger than the other bursts which all appear to occur in eclipse
ingress or egress.  This is not seen, indicating that the burst
reprocessing is probably dominated by the disk.

\subsection{Optical flares in quiescent systems}

\citet{Hakala:1998a} suggested that UW~CrB might be a quiescent LMXB,
possibly harboring a black hole.  In this case an alternative flaring
mechanism is needed.  Quiescent LMXBs containing both black holes and
neutron stars do undergo relatively rapid optical flaring (e.g.,
\citealt{Zurita:2003a}; \citealt{Hynes:2003a}), and this behavior was
already considered by \citet{Jonker:2003a} as a possible explanation
for the X-ray flare of \citet{Mukai:2001a}.  Indeed, the
black hole candidate XTE~J1650--500 has exhibited {\em non-thermal}
X-ray flares otherwise very similar to type-I bursts in a low state
\citep{Tomsick:2003a}.  In the case of
A\,0620--00 at least, the optical events can be relatively rapid, occurring
on timescales comparable to those seen here \citep{Hynes:2003a}.
While the observed amplitudes are much smaller, that is probably a
consequence of dilution of the accretion light by the companion star,
and the undiluted disk light could undergo variations of a factor of
two or more.  In spite of these similarities, the lightcurves shown in
Fig.~\ref{OrbitFig} appear completely different to those of quiescent
LMXBs.  The burst-like events we see are single discrete events, and
do not appear to belong to the general range of flickering behavior.
The similarity of burst fluences is also rather unlike the kind of
stochastic events seen in quiescent LMXBs to date.  We therefore
consider this explanation of the bursts to be very unlikely, and that
the type-I burst interpretation is far more plausible.

\subsection{The nature of UW~CrB}

Now that we have seen both an X-ray burst-like event
\citep{Mukai:2001a}, and multiple optical events, it is hard to escape
the conclusion that UW~CrB is an X-ray burster.  This immediately
rules out white dwarf and black hole models for the system; the
compact object is a neutron star.  The burst properties and recurrence
time further constrain the accretion rate.  The short burst recurrence
time rules out a quiescent system.  Such systems can exhibit type-I
bursts, but the recurrence time is estimated to be 10--60\,yrs, or
even more \citep{Cornelisse:2002a}, a direct consequence of the very
low accretion rate onto the neutron star.

The probable 4--12\,hr recurrence time and 10--20\,s burst duration
are both typical of intermediate luminosity bursting LMXBs
(\citealt{vanParadijs:1988a}; \citealt{Strohmayer:2003a}) and suggest
that UW~CrB is not accreting at an unusually low rate; hence its low
X-ray brightness must indicate either extreme distance or an ADC
source as discussed in Section~\ref{IntroSection}.  In contrast, for
example, Aql X-1 was observed to exhibit optical bursts lasting over a
minute and recurring once per hour \citep{Robinson:1997a}.  Both short
recurrence times and long bursts are believed to be often associated
with low accretion rates \citep{Strohmayer:2003a}, yet are not seen in
UW~CrB.

It is also of value to compare our observations with those of
GS~1826--24 \citep{Homer:1998a}.  Both GS~1826--24 and UW~CrB are
neutron star LMXBs having orbital periods $\sim2$\,hr; thus the sizes
of the two systems should be very similar.  If we assume that the
intrinsic optical burst luminosity is simply a function of the X-ray
burst strength and reprocessing disk area (and hence orbital period),
then we might expect similar luminosity reprocessed bursts in the two
systems, assuming that X-ray bursts have comparable peak luminosities,
and similar disk geometries.  GS~1826--24 is a low-inclination system,
however, whereas UW~CrB is eclipsing and hence high inclination.  In a
high inclination system we see less projected disk area, so if
reprocessing is dominated by the disk (as suggested by the lack of
phase dependence in the optical burst fluences), then we would expect
to observe weaker optical bursts at a given distance.  This is not the
case, however; both UW~CrB and GS~1826--24 are of comparable optical
brightness and exhibit optical bursts rising to a peak $\sim2\times$
the persistent optical level, indicating that the observed optical
burst flux is actually comparable in the two systems.  This would
imply, subject to the assumptions made, that UW~CrB is somewhat closer
than GS~1826--24, or at least not much further away.  The distance to
GS~1826--24 is estimated at less than $7.5\pm0.5$\,kpc by
\citet{Kong:2000a}, so unless bursts in UW~CrB are substantially more
luminous than in GS~1826--24, or reprocessing into optical emission is
much more efficient, we would expect a comparably nearby distance for
UW~CrB.  This contradicts the extreme distances inferred for the
dipper scenario (e.g., 75\,kpc \citealt{Mukai:2001a}) and instead
favors the ADC model.  Note that this argument is essentially similar
to that made by \citet{Jonker:2003a}, but complements it as we have
used the optical burst brightness rather than the persistent level.
The argument is approximate and model dependent, but is sufficient to
indicate that UW~CrB should not be $10\times$ further away than
GS~1826--24 as has been previously suggested.

\section{Conclusions}
We have reported the discovery of several resolved optical bursts from
UW~CrB.  These are almost certainly reprocessed type-I X-ray bursts,
clarifying several characteristics of the source.  i) For type-I
bursts, the compact object must be a neutron star rather than a black
hole or white dwarf.  ii) The burst rate is relatively high,
indicating an active rather than quiescent system, and thus a distance
greater than a few kpc. iii) The optical burst flux is comparable to
the similar source GS~1826--24, suggesting a comparable distance ($\la
10$\,kpc).  It is thus most likely that UW~CrB is an ADC source (as
also argued by \citealt{Jonker:2003a}) rather than a distant dipper.
Given its high Galactic latitude and intermediate distance, however,
it must still be situated in the Galactic halo.
\acknowledgments
RIH is supported by NASA through Hubble Fellowship grant
\#HF-01150.01-A awarded by the Space Telescope Science Institute,
which is operated by the Association of Universities for Research in
Astronomy, Inc., for NASA, under contract NAS 5-26555.  We are
indebted to Ed Nather and Anjum Mukadam for making Argos available to
the McDonald community and providing considerable assistance in its
use, and also grateful to Brad Schaefer for discussions and
constructive skepticism on the first of the bursts.  This work has
made use of the NASA ADS Abstract Service.

\clearpage



\begin{thebibliography}{}

\bibitem[Cornelisse et al.(2002)]{Cornelisse:2002a} Cornelisse, R., 
Verbunt, F., in't Zand, J.~J.~M., Kuulkers, E., \& Heise, J.\ 2002, \aap, 
392, 931 

\bibitem[Ergma \& Vilhu(1993)]{Ergma:1993a} Ergma, E., Vilhu, O.\
  1993, \aap, 277, 483

\bibitem[Gioia et al.(1990)]{Gioia:1990a} Gioia, I.~M., Maccacaro, 
T., Schild, R.~E., Wolter, A., Stocke, J.~T., Morris, S.~L., \& Henry, 
J.~P.\ 1990, \apjs, 72, 567 

\bibitem[Gottwald et al.(1986)]{Gottwald:1986a} 
Gottwald, M., Haberl, F., Parmar, A.~N., \& White, N.~E.\ 1986, \apj, 308, 
213 

\bibitem[Grindlay et al.(1978)]{Grindlay:1978a} Grindlay, J.~E., 
McClintock, J.~E., Canizares, C.~R., Cominsky, L., Li, F.~K., Lewin, 
W.~H.~G., \& van Paradijs, J.\ 1978, \nat, 274, 567 

\bibitem[Hakala et al.(1998)]{Hakala:1998a} Hakala, P.~J., Chaytor, 
D.~H., Vilhu, O., Piirola, V., Morris, S.~L., \& Muhli, P.\ 1998, \aap, 
333, 540 

\bibitem[Homer, Charles, \& O'Donoghue(1998)]{Homer:1998a} Homer, 
L., Charles, P.~A., \& O'Donoghue, D.\ 1998, \mnras, 298, 497 

\bibitem[Hynes et al.(2003)]{Hynes:2003a} Hynes, R.~I., Charles, 
P.~A., Casares, J., Haswell, C.~A., Zurita, C., \& Shahbaz, T.\ 2003, 
\mnras, 340, 447

\bibitem[Hynes et al.(2004)]{Hynes:2004a} Hynes, R.~I., et al., in
  preparation 

\bibitem[Jonker et al.(2003)]{Jonker:2003a} Jonker, P.~G., van der 
Klis, M., Kouveliotou, C., M{\' e}ndez, M., Lewin, W.~H.~G., \& Belloni, 
T.\ 2003, \mnras, 346, 684 

\bibitem[Kong et al.(2000)]{Kong:2000a} Kong, A.~K.~H., Homer, L., 
Kuulkers, E., Charles, P.~A., \& Smale, A.~P.\ 2000, \mnras, 311, 405 

\bibitem[Morris et al.(1990)]{Morris:1990a} Morris, S.~L., Liebert, 
J., Stocke, J.~T., Gioia, I.~M., Schild, R.~E., \& Wolter, A.\ 1990, \apj, 
365, 686 

\bibitem[Mukai et al.(2001)]{Mukai:2001a} Mukai, K., Smale, A.~P., 
Stahle, C.~K., Schlegel, E.~M., \& Wijnands, R.\ 2001, \apj, 561, 938 

\bibitem[Nather \& Mukadam(2004)]{Nather:2004a} Nather, R.~E.~\& 
Mukadam, A.~S.\ 2004, \apj, 605, 846 

\bibitem[Robinson \& Young(1997)]{Robinson:1997a} Robinson, E.~L.~\& 
Young, P.\ 1997, \apjl, 491, L89 

\bibitem[Schoembs \& Zoeschinger(1990)]{Schoembs:1990a} Schoembs, 
R.~\& Zoeschinger, G.\ 1990, \aap, 227, 105 

\bibitem[Strohmayer \& Bildsten(2003)]{Strohmayer:2003a} Strohmayer,
  T., Bildsten, L.\ 2003, to appear in Compact Stellar X-Ray Sources,
  eds.\ W.~H.~G. Lewin and M. van der Klis, Cambridge University Press

\bibitem[Tomsick, Kalemci, Corbel, \& Kaaret(2003)]{Tomsick:2003a} 
Tomsick, J.~A., Kalemci, E., Corbel, S., \& Kaaret, P.\ 2003, \apj, 592, 
1100 

\bibitem[van Paradijs, Penninx, \& Lewin(1988)]{vanParadijs:1988a} van 
Paradijs, J., Penninx, W., \& Lewin, W.~H.~G.\ 1988, \mnras, 233, 437 

\bibitem[Zurita, Casares, \& Shahbaz(2003)]{Zurita:2003a} Zurita, 
C., Casares, J., \& Shahbaz, T.\ 2003, \apj, 582, 369 

\end{thebibliography}
\end{document}